# Unpredictability and Computational Irreducibility

HERVE ZWIRN [†] and JEAN-PAUL DELAHAYE [‡]

[†] *UFR de Physique (Université Paris 7), CMLA (ENS Cachan) & IHPST(CNRS). herve.zwirn@m4x.org*

[‡] *Laboratoire d'Informatique Fondamentale de Lille (CNRS). jean-paul.delahaye@lifl.fr*

We explore several concepts for analyzing the intuitive notion of computational irreducibility and we propose a robust formal definition, first in the field of cellular automata and then in the general field of any computable function $f$ from $\mathbf{N}$ to $\mathbf{N}$. We prove that, through a robust definition of what means "to be unable to compute the $n^{th}$ step without having to follow the same path than simulating the automaton or to be unable to compute $f(n)$ without having to compute $f(i)$ for $i=1$ to $n$-1", this implies genuinely, as intuitively expected, that if the behavior of an object is computationally irreducible, no computation of its $n^{th}$ state can be faster than the simulation itself.

## 1. Introduction

It is now common knowledge that it is not because the behavior of a system is deterministic that it is possible to predict it. That has been proven in mathematics in the first part of the XX[th] century through the work of Kurt Gödel on the formal axiom systems and of Alan Turing[1] on the computing machines. That has also been proven in physics after Henri Poincare, Edward Lorenz and the subsequent works on the so called "deterministic chaos" during the second half of the XX[th] century[2]. So, we now know that even if a system is deterministic it can be the case that we can't predict its behavior in the long run. The reasons for this unpredictability can be multiple. If the behavior of a dynamical non linear system is not predictable, it is because very near initial conditions can lead to very far states after a certain time. So, predicting the behavior in the long term would mean knowing initial conditions with an infinite precision which is impossible. As far as computing machines are concerned, the halting problem is known to be undecidable. That means that no algorithm, if fed with the description of a particular Turing machine, can tell if this machine is going to stop or to compute forever. Incidentally, a consequence of this undecidability is that for every formal system, there exists a Turing machine that doesn't halt but for which it's impossible to prove that it will never stop. In this case, what we can't predict is a fact about a fully deterministic machine.

---

[1] [Chaitin, 1992, 2001], [Calude, 2002], [Zwirn, 2000].

[2] [Lorenz, 1993], [Zak, 1997], [Zwirn, 2000], [Ott, 2002], [Schuster, 2005], [Bishop, 2008].

Another kind of unpredictability is of interest. It concerns also computing machines but is linked with their computation time. Predicting the behavior of a computation machine is to be able to find the result it computes faster than the machine itself. Of course in any case, using a modern computer obviously enables to get faster the result of a given program running on an older one. A more interesting definition is: given a Turing machine[3] computing a definite function, is there another Turing machine computing the same function faster (i.e. in a smaller number of steps)? It is useful to be more precise and to restrict the type of the Turing machines allowed to compete. The efficiency of different kinds of Turing machines is not the same. For example, when deciding whether a string of length $l$ is a palindrome, a 1-tape Turing machine will need a number of steps of $O(l^2)$ whereas a 2-tape Turing machine will use only $O(l)$[4]. Therefore an immediate question occurs: Is it possible by adding an arbitrary number of tapes to increase indefinitely the gap between the speed of 1-tapeTuring machines and the speed of $k$-tape Turing machines? The answer is no for a well known theorem states that "given any $k$-tape Turing machine M operating within time $f(n)$, we can construct a 1-tape Turing machine M' operating within time $O(f(n)^2)$ and such that for any input $x$, $M'(x) = M(x)$[5]". That means that whatever fast $k$-tape machine is considered for computing a function, there is a 1-tape Turing machine doing the same job in just the square of the time needed by the $k$-tape machine. So the best we can do is achieving quadratic savings in time. The link with our subject of unpredictability is clear: if a computation is performed through the fastest algorithm that can perform it, it will not be possible to predict the result. Of course, these considerations are of interest only when infinite processes are considered (i.e. computation of a function $f(n)$ for all values of $n$). For a finite computation of a given data $d$ the fastest algorithm that computes $d$ is always something like: print "$d$". A more precise definition will be given below and more will be said as well about the so called "speed-up" theorems.

The question of unpredictability can be put forward in a more direct way: given a physical system whose behavior can be calculated by simulating explicitly each step in its evolution, is it always possible to predict the outcome without tracing each step? That means: is there always a shortcut to go directly to the $n^{th}$ step? A computer is of course such a physical system. Wolfram conjectured[6] that in most cases the answer is no. This "no" expresses the concept of computational irreducibility (CIR). This question has been widely analyzed in the context of cellular automata (CA) by Wolfram[7]. A cellular automaton (CA) is computationally irreducible if in order to know the state of the system after $n$ steps there is no other way than to evolve the system $n$ times according to the equations of motion. That means that there is no short-cut, no way to compress the dynamic. Thus the system appears to be unpredictable. The intuition behind this definition is that there is no other way to reach the $n^{th}$ state than to go through the ($n$-1) previous ones. The consequence is that it's impossible to reach the $n^{th}$ state faster than the automaton itself. While really appealing from the intuitive standpoint, this definition lacks for robustness. If we take it at face value, asking that there is no other possibility than to go through exactly the same states than the automaton itself, it's too restrictive. Imagine for example, that there is an algorithm $A_1$ such that it gives for each input $n$ a result differing only by a small difference from the $n^{th}$ state of the automaton and that for computing the result for $n$, it goes through all the previous results for $i < n$. Imagine as well that there is an algorithm $A_2$ that computes the $n^{th}$ state of the automaton from the $n^{th}$ result of the algorithm $A_1$. Then, the composed algorithm $A_2oA_1$ (where the symbol o stands for the composition) will give the $n^{th}$ state of the automaton when given $n$ as input but will not follow the same path than the automaton itself. Nevertheless, if no other algorithm is able to give directly the $n^{th}$ state of the automaton, we will not be willing to consider the existence of $A_2oA_1$ as a counter example of the fact that the automaton is CIR. That means that demanding that the only way to reach the $n^{th}$ state is to follow the exact path of the automaton through the (*n*-1) previous states is too restrictive. So a first question is: how far can we depart from this path? A second question is related to the computation time. Could it be enough to say that a CA is CIR if it's impossible to know its $n^{th}$ state faster than itself even if it is possible to compute this $n^{th}$ state without having to go through all the previous states?

---

[3] [Garey, 1979], [Hopcroft, 1979], [Hartley, 1987], [Odifreddy, 1989], [Papadimitriou, 1994] [Wolfram, 2002], [Goldreich, 2008].
[4] [Papadimitriou, 1994] p.30 or [Goldreich, 2008] p.33.
[5] [Papadimitriou, 1994] p.30.
[6] [Wolfram, 1985]. See also [Israeli, 2004], [Zenil, 2011].
[7] [Wolfram, 2002].

Even if these questions have been raised inside the cellular automata field, they concern as well the field of all computable functions. What does that mean for a function *f(n)* from **N** to **N** (the set of all positive or null integers) to be CIR? Is there any function *f(n)* such that it's necessary to compute all the *f(i)* for *i* = 1 to *n*-1, to get *f(n)*?

In this paper, we deal with these questions and attempt to find a robust formal definition of the concept of computational irreducibility. Our goal is general and we look for a concept of computational irreducibility that is applicable to any system. Computational irreducibility is related to a system and a computation model. For the sake of simplicity, we'll use as our typical examples of systems the cellular automata and we'll consider Turing machines as implementing the concept of computation. The reader will easily convince himself that these choices don't imply any loss of generality and at the end of the paper, we'll aim at giving the most general form to our results so as not to be restricted to the field of cellular automata which is used only as a suitable example.

## 2. Turing machines

We assume the reader to be familiar with the concept of Turing machines[8] but we give nevertheless the basic notions.

A Turing machine is a theoretical device capable of manipulating a list of cells called tape (infinite in principle), using an access pointer called the head through a finite program. Each cell can contain a symbol from a finite alphabet. The time is discrete and each instruction is executed in one step of time. The head is always positioned over a particular cell which it is said to scan. The machine can be in internal states (members of a finite set always containing at least a start state and a halting state). At time 0, the head is supposed to scan the leftmost cell (the string written on the tape at the beginning is considered as input, possibly empty) and the machine to be in the start state. At each step, the head reads the symbol written on the scanned cell, erases it or writes another symbol, goes left or right and changes its internal state. The program of the machine is a finite rule (sometimes called the transition function) which states exactly what the head must do depending on the symbol written on the scanned cell and the internal state of the machine. When the machine reaches the halting state the computation is finished and what is written on the tape is the result of the computation. Of course, the machine can possibly never stop. It is also possible to consider Turing machines with several tapes. As we'll see, the consideration of multi-tape Turing machines is important when dealing with the concept of complexity. Multi-tape Turing machines are also very useful to facilitate the design of machines that compute functions of interest.

The main point with the Turing machines model is that it is very simple and that through the Church-Turing thesis[9], it allows the computation of any computable function. More precisely, the Church-

---
[8] [Garey, 1979], [Hopcroft, 1979], [Hartley, 1987], [Odifreddy, 1989], [Papadimitriou, 1994] [Wolfram, 2002], [Goldreich, 2008].

Turing thesis says that a function can be computed by some Turing machine if and only if it can be computed by some machine of any other reasonable and general model of computation. Put simply, that means that a function is in any way computable if and only if it can be computed with a Turing machine.

The fact that the description of any Turing machine through its transition function is itself computable shows that there exist Turing machines able to simulate any other Turing machine. Such a machine is called a Universal Turing machine. A Universal Turing machine U is such that given the index *i* of any Turing machine T under a given numbering, it simulates the computation of T on every input argument *m*:

For all *i*, if T is the *i*[th] Turing machine, then for all *m*, U(*i*,*m*) = T(*m*).

We'll see that Universal Turing machines are fundamental tools for defining many important notions in the following.

## 3. Some general speed-up results

**Notations**
We'll need in the following some notations for comparing the order of magnitude of different functions. We recall here the standard notations.
  i)   $f(n) = O(g(n))$ if there are constants $c > 0$, $n_0 > 0$ such that $\forall n > n_0$, $|f(n)| \leq c|g(n)|$.
  ii)  $f(n) = o(g(n))$ if $\lim_{n \to \infty} f(n)/g(n) = 0$.
  iii) $f = \omega(g)$ if there is a constant $n_0 > 0$ such that $\forall c > 0$, $|f(n)| > c|g(n)|$.
  iv)  $f(n) = \Omega(g(n))$ if there is a constant $c > 0$ such that $|f(n)| \geq c|g(n)|$ infinitely often.

It is well known that the time complexity of a problem may depend on the model of computation. We mentioned above the problem of deciding if a string is a palindrome which is $O(n^2)$ in the 1-tape Turing machines model and $O(n)$ in the 2-tape Turing machines model.

If we adopt the computation model of the *k*-tape Turing machines, some results help understanding the limits on the savings that can be expected either by increasing the number of tapes or by designing more efficient machines doing the same computation.

A first result[10] says that we can't expect more than a quadratic saving through allowing an arbitrary number of tapes.

**Theorem 3.1.** *Given any k-tape Turing machine* M *operating within time f(n), it's possible to construct a* 1*-tape Turing machine* M' *operating within time* $O(f(n)^2)$ *and such that for any input x,* M(*x*)=M'(*x*).

The meaning of this result is that the best *k*-tape machine that can be designed for doing a computation will never operate in less that $O(f(n)^{1/2})$ if the best 1-tape Turing machine doing the same computation operates in a time $O(f(n))$.

---

[9] [Gandy, 1980], [Copeland, 2002].
[10] [Papadimetriou, 1994] p.30.

More generally, the Cobham-Edmonds thesis[11] states that a problem has time complexity $t$ in some "reasonable and general" model of computation if and only if it has time complexity poly($t$) in the model of 1-tape Turing machines[12]. The time complexity of the problems in all reasonable models is polynomially related.

A second result[13] is known as linear speed-up.

**Theorem 3.2.** *For any k-tapes Turing machine* M *operating in time f(n) there exists a k'-tapes Turing machine* M' *operating in time f'(n)=εf(n)+n* (*where ε is an arbitrary small positive constant*) *which simulates* M.

This linear speed-up means that the main aspect of complexity is captured through the function $f(n)$ irrespectively of any multiplicative constant. **DTIME**($f(n)$) is the class of functions[14] computable by a $k$-tape Turing machine in $f(n)$ steps. This result means that **DTIME**($f(n)$) = **DTIME**($\varepsilon f(n)$) and so it's legitimate to define **DTIME**($f(n)$) as the class of functions computable by a Turing machine in O($f(n)$) steps.

More interesting is the Time Hierarchy theorem[15] which states that for 2-tape Turing machines:

**Theorem 3.3. DTIME**($f(n)$) *is strictly included in* **DTIME** ($f(n)g(n)$) *when* $g(n) = \omega(\log n)$ *and* $f(n) > n$.

For example, there are some functions that are computable in O($n^2 \log^2 n$) and not computable in O($n^2$). Using the Cobham-Edmonds thesis this gives similar hierarchy theorems for any reasonable models of computation.

A simple consequence is theorem 3.4:

**Theorem 3.4.** *A universal Turing machine cannot be significantly sped-up* (*more than by a factor* O($\log^2(n)$)).

**Proof**: The reason why such a universal Turing machine cannot be significantly sped-up is the following. A multi-tape universal Turing machine needs only be slower by logarithmic factor[16] compared to the machines it simulates. Assume T computes in time O($t(n)$), then U($i$,.) computes in time O($t(n) \log n$). If U could be significantly sped-up then any function simulated by U would as well be sped-up the same way, up to the factor $\log n$. Then, let's consider a function $f$ in **DTIME**(O($n^2 \log^2 n$)) - **DTIME**(O($n^2$)). Speeding-up U by a factor O($\log^2(n)$) would mean being able to compute $f$ through U in a time O($n^2$), which is impossible.

---

[11] [Cobham, 1965], [Goldreich, 2008] p.33.
[12] Poly(t) stands for any function polynomial in t.
[13] [Papadimetriou, 1994] p.32.
[14] More precisely the class of decision problems.
[15] [Goldreich, 2008] p.130.
[16] [Goldreich, 2008] p.134.

# 4. Chaitin-Kolmogorov complexity and logical depth of Bennett

We give here the basic notions of algorithmic complexity and logical depth. A good reference is the book by Li and Vitanyi[17]. We'll then explore the possible conceptual links between these two notions and CIR.

## 4.1. The Chaitin-Kolmogorov complexity

Intuitively, the Chaitin-Kolmogorov complexity[18] (sometimes called simply algorithmic complexity) K($s$) of a string s is the length of the shortest computer program $p$ that outputs $s$ if it is given as input to a universal machine U: K($s$) = min{$\ell(p)$ | U($p$) = $s$} where $\ell(p)$ is the length of $p$.
This program is called the minimal program or the shortest description for $s$ and sometimes noted $s^*$. It is possible to show that this definition depends on the choice of the universal machine only up to a constant. More precisely:

**Theorem 4.1.1.** (*invariance theorem*) *Given two Universal Turing machines* U *and* V, *there exists a constant* $C_{UV}$ *depending only of* U *and* V *such that for every s:* $|K_U(s) - K_V(s)| < C_{UV}$.

**Proof:** Roughly, this can be understood through the fact that the universal Turing machine U can be simulated through the other one V by a simulation program $p_{VU}$. So, if $s^*_U$ is the minimal program for $s$ relatively to U, $p_{VU}$ o $s^*_U$ (where o stands here for composition) is a program computing $s$ on V. Hence $K_V(s) \leq \ell(p_{VU} \text{ o } s^*_U) = \ell(p_{VU}) + \ell(s^*_U) = K_U(s) + \ell(p_{VU})$ (and vice versa).

This invariance theorem gives all its interest to the definition because it states that the algorithmic complexity of a string is a good measure (choosing a universal Turing machine is roughly similar to choosing the zero of a temperature scale for a thermometer). It also shows that from an asymptotic point of view, the complexity of a string does not depend on the chosen machine (the simulation program becomes negligible).

The algorithmic complexity of a string $s$ is a measure of how regular is the string. If the string contains many redundancies it will be easy to compress and its complexity will be low. For instance, the string $(01)^{1000}$ of one thousand times "01" is easy to describe in a way much shorter than its length (we just did it). On the contrary, if there is no redundancy, the only way to describe a random string is to enumerate all of its bits. So the shortest program with the output $s$ will be "print $s$" and its length will be of the order of magnitude of the length of $s$ (plus the length of the program print). A finite string will be random (i.e. with no redundancy at all) if its complexity is roughly equal to its size. Equivalently, that means that it is not compressible. The right statistical definition of a random infinite string has been given by Martin-Löf[19][20]. An interesting point is the link between the algorithmic complexity and randomness for infinite strings.

**Theorem 4.1.2.** (*infinite binary sequence*) *An infinite binary sequence s is random iff there is a constant c such that for all n:* K($s_{1:n}$) ≥ n – c (*where $s_{1:n}$ stands for the initial segment of the n first bits of s*)

---

[17] [Kolmogorov, 1965], [Chaitin, 1992, 2001], [Li, 1997], [Calude, 2002], [Zenil, 2011, 2012].
[18] We present here the so called "prefix complexity" in which no program is a proper prefix of another program. There are many technical reasons for imposing this restriction. See [Li, 1997] for an extensive presentation.
[19] [Martin-Löf, 1966]. In precise terms $s$ is random if it is not contained in any $G_\delta$ set determined by a constructive null cover.
[20] [Delahaye, 1993, 1998], [Chaitin, 2001], [Calude, 2002], [Downey, 2010].

Another point worth noticing is the fact that among the $2^n$ binary strings of length $n$, less than a proportion of $2^{-p}$ have a complexity smaller than $n-p$. The reason why is easy to understand. There are less than $2^k$ strings of length inferior to $k$, then less than $2^k$ programs of length inferior to $k$. So there is a maximum number $2^k$ of strings of complexity smaller than $k$. Then the proportion of strings of complexity smaller than $k$ among the $2^n$ strings of length $n$ is less than $2^{k-n}$. Equivalently (by letting $k = n-p$) less than a proportion of $2^{-p}$ have a complexity smaller than $n-p$. That means that almost all strings are incompressible and hence random.

### 4.2. Bennett's logical depth

Bennett's logical depth[21] is an attempt to measure the amount of non random or "useful" information in a string. Roughly, the logical depth of a string $s$ is defined as the time required by a universal Turing machine to generate $s$ from its shortest description[22]. Note that this time is at least equal to the length of the string $s$ since $s$ has to be written (which needs at least as many steps as the number of bits of s).

The computation models used to estimate the time of computation need to be reasonable. Bennett considers what he calls "fast universal Turing machines". For example, these machines must be able to run a "print s" program in a time linear in the length of $s$. More generally, a fast universal Turing machine must be able to simulate any computation done on another machine in a time bounded by a polynomial linear in the time needed by the other machine, whatever other machine is considered.

Intuitively, an object is deep if it contains some redundancy (hence is not random), but an algorithm requires extensive resources to exploit that redundancy.

It's also possible to define the depth of a string $s$ relative to another string $w$, which is the computation time to produce $s$ from $w$ by the minimal program of $s$. The more the depth D($s$) of a string $s$ will be large relative to its length $n$, the deeper $s$ will be. For example a string $s$ with D($s$)=O($n^n$) will be deeper than a string $s$' with D($s$')=O($n^2$). Some strings are supposed to have a non linear depth in this precise meaning. That's the case of the string composed of the $n$ bits ($a_1, a_2, \ldots, a_n$) where $a_i$ is 1 if the $i^{th}$ Turing machine (under a given numbering) halts and 0 otherwise. A problem with the logical depth is that it is not computable (there is the same problem with the algorithmic complexity). As a result it's often impossible to prove things rigorously. A large literature has been extensively written these last years on the subject and some authors[23] have proposed more sophisticated but computable definitions of depth.

In this paper, we are mainly interested about the possible conceptual links between logical depth and CIR. Hence, we will not dive into technical subtleties out of the scope of our subject and will adopt the simplest (if not the most correct) definition of the logical depth: D($s$) is the computation time of the minimal program for $s$.

The main point about the logical depth is that it can easily be seen that it is a measure of the amount of useful information while the algorithmic complexity is clearly a measure of random information.

---

[21] [Bennett, 1988, 2003, [Antunes, 2009], [Ay, 2010]].
[22] Actually, for technical reasons, this definition is not totally satisfying and the correct one is: the depth at significance level l of a string s is the least time required by a universal Turing machine to generate s by a program that is not compressible itself by more than l bits. We'll ignore this subtlety in the following.
[23] See for example [Moser, 2008].

A random string with maximal algorithmic complexity (of order of magnitude of its length) is not deep. Logical depth and algorithmic complexity are complementary notions. A deep string can be seen as a highly organized object or as an object produced by a long computation.

A string with a simple organization (*n* times "0" for instance) can be deep if *n* is deep but will always have a low depth relative to its length. For a string to be really deep, it will necessarily be produced after a long computation having let some tracks in its structure. The slow growth law[24] expresses the fact that it is very unlikely that a deterministic program transforms quickly a shallow string into a deep one. This is an indication that the logical depth concept is reaching its goal of describing the organization hidden in a string and that it is a very important mathematical notion that can even be used in concrete applications[25].

## 5. Elementary cellular automata

For the sake of simplicity and as a useful intuitive guide but without any loss of generality, we will address the problem of defining the computational irreducibility first inside the framework of Elementary Cellular Automata (ECA). Cellular automata (CA) were originally introduced by von Neumann and Ulam[26] in the 1940's as a possible way of simulating self reproduction in biological systems. A CA is a dynamical system composed of a lattice of cells inside a one or many dimensions array. Each cell can contain a value from a given finite alphabet. The system evolves in time according to an update rule that gives a cell's new state as a function of the values of the other cells in a given neighborhood (for instance the eight immediate neighbors in a square array). A ECA is a one dimensional CA which has two possible values for each cell (0 or 1) and update rules that depend only on the two nearest neighbor values. According to Wolfram[27]:

*" the evolution of an elementary cellular automaton can completely be described by a table specifying the state a given cell will have in the next generation based on the value of the cell to its left, the value the cell itself, and the value of the cell to its right. Since there are $2 \times 2 \times 2 = 2^3 = 8$ possible binary states for the three cells neighboring a given cell, there are a total of $2^8 = 256$ elementary cellular automata, each of which can be indexed with an 8-bit binary number".*

A point worth noticing is that the number of update rules to perform to reach the $n^{th}$ configuration going successively through all the configurations is growing at most as $n^2$. That's easy to see through considering an initial state with only one black cell. Then computing the next configuration needs applying one of the 8 rules describing the automaton. At the next step, there can be at most 3 black cells, then 5 black cells and so on. If we number 0 the initial configuration, the $n^{th}$ configuration contains at most $2n+1$ black cells. Thus, provided that no intermediate configuration collapses, the number of rules to apply to reach the $n^{th}$ configuration is: $3+5+\ldots+(2n-1) = n^2-1$. If the length of the initial configuration is *l*, then the number of update rules to perform will be *l*+2 for the next configuration, then *l*+4, and so on… The total number of update rules to perform will be at most $n^2+n(l-1)-1$.

---

[24] [Bennett, 1988].
[25] [Zenil, 2011].
[26] [von Neumann, 1966] [Ulam 1952].
[27] [Wolfram, 1983], [Wolfram, 2002).

Does that mean that the computation of the $n^{th}$ configuration of every ECA among the 256 possible ones always needs to perform $n^2$ rules? No, since some ECA's show trivial behavior[28]. Rules 0, 40 or 96 give immediately vanishing configurations. Rules 4,12, 36 or 76 give stables configurations with a unique black cell under the initial one. These ECA can be simulated[29] in time O($n$) and of course the computation of the $n^{th}$ configuration can be done in a constant time since all the configurations are identical. A little bit more interesting are the rules 2, 6, 16, and many similar others giving rise to a sloping black line of one cell length. The computation of the $n^{th}$ configuration can be done directly in time O($n$). The simulation can be done either in time O($n^2$) or time O($n$) depending on the fact that we demand that all the successive configurations keep written or not. In the following we'll just demand that all the configurations appear and not that they keep written. So, the simulation can be done in O($n$).

Consider now the ECA rule 158 (see figure 1). It's easy to see that the $n^{th}$ configuration is always a string of length $2n+1$ with the following structure: for $n$ odd, 1110011001100..…0011 and for $n$ even, 111011101110..…11101. So, while simulating rule 158 needs a time O($n^2$), it is not complicated to compute directly the $n^{th}$ configuration in a time O($n$).

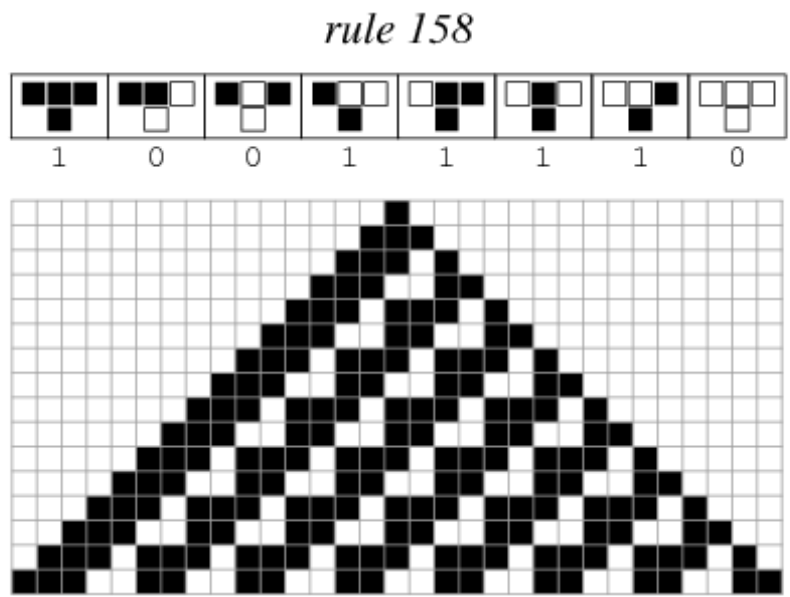

Figure 1.

The same situation appears with many other ECA. Consider the ECA rule 90 (see figure 2). The situation is simple since the $2^n$ th configuration is the string "$(10)^n 1$". The configuration $2^n+1$ is the string "$1(0)^{2n-1}1$". Here again, while the simulation of rule 90 needs a time O($n^2$), it's easy to compute directly the $n$th configuration in time O($n$).

---

[28] In the following, we'll consider only the behavior of ECA from an initial state with only one black cell.
[29] A simulation of a ECA A is the enumeration of the successive configurations of A. We'll define below a ECA Turing machine representing a ECA A as a Turing machine computing successively all the configurations of A.

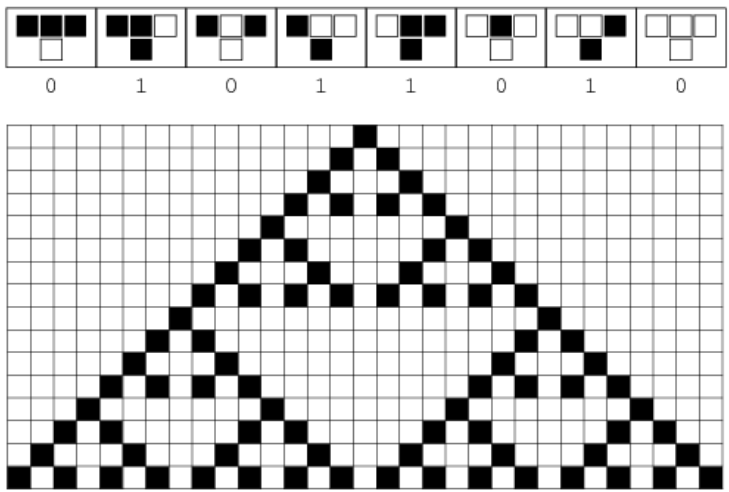

Figure 2.

For all these automata (rules 2, 6, 16, … giving only a slopping black line, rules 90, 158, … giving more complex but regular configurations) computing the $n^{th}$ configuration can be done in time $O(n)$.

However this is not always the case. Consider for example the ECA rule 30 (see figure 3).

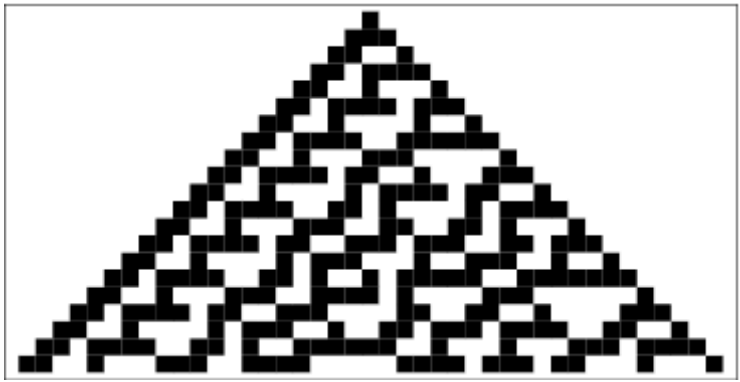

Figure 3.

It seems much more difficult to see any reasonable way to find a rule giving directly the structure of the $n^{th}$ configuration. That seems as well difficult with ECA rule 110 (see figure 4).

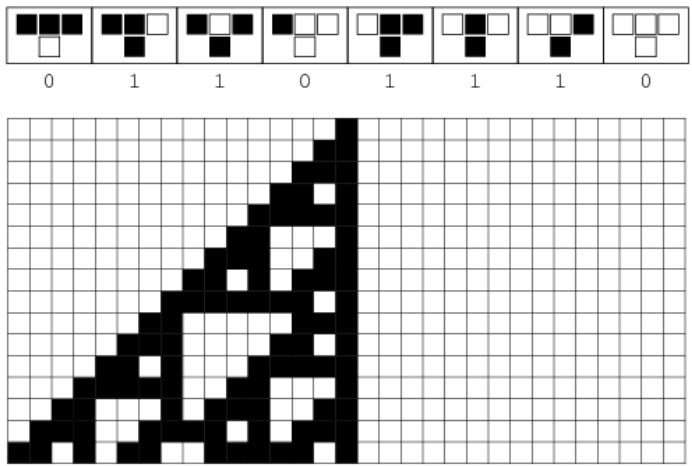

Figure 4.

That looks even more difficult if we have a look at a larger number of configurations (see figure 5).

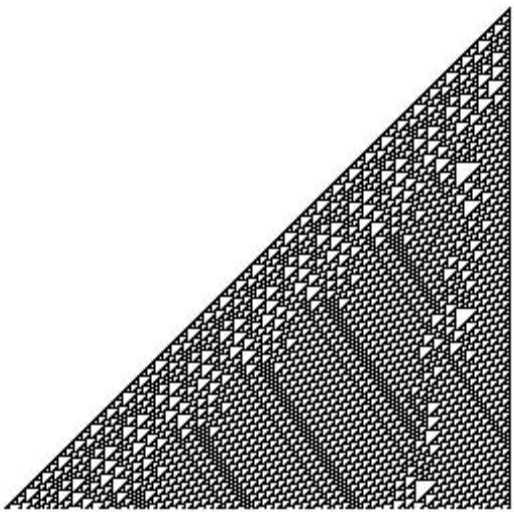

Figure 5.

Rule 110 is of particular interest since it has been proved by Cook[30] to be capable of universal computation. It's up to now the simplest rule known to be universal.

We can draw a classification[31] according to the following classes (noticing that the simulation time must be at least O($n$) since a simulation is the enumeration of $n$ successive configurations and that the direct computation time can't be greater than the simulation time):

---

[30] [Cook 2004].
[31] Be careful not to confuse this classification with Wolfram's classification which bears some similarities with this one.

| ECA | simulation time | $E_n$ direct computation time | Class |
|---|---|---|---|
| Rules, 0, 8, 32, 40, 96, ... and Rules 4, 12, 36, 44, 76 ... | $O(n)$ | $O(1)$ | 1 |
| Rules 2, 6, 16, 24, ... | $O(n)$ | $O(n)$ | 2 |
| Rules 18, 26, 90, 158, ... | $O(n^2)$ | $O(n)$ | 3 |
| Rules 90, 110, ... ?? | $O(n^2)$ | ? | 4 |

Class 2 and 4 are the only possible classes for CIR ECA. Automata in class 2 have the property that the number of cells that change between two successive configurations is bounded by a constant and so is the time to compute the $(n+1)^{th}$ configuration from the $n^{th}$ configuration. We don't want to consider them as CIR. Indeed, their behavior is very simple and hence can easily be predictable. So the only possible class for CIR ECA is class 4. The consideration of such automata naturally raises the question of whether it is possible - but just not obvious - to find a rule for directly computing the $n^{th}$ configuration of such automata or if this is really impossible as claimed by Wolfram. So the question is: "is there any automaton in class 4?".

This question can actually be split into:

1) is it possible to compute the $n^{th}$ configuration without having to perform $n^2$ rules?
2) is it possible to compute the $n^{th}$ configuration without computing the ($n$-1) previous ones?

An answer "no" to the second question implies an answer "no" to the first one if all the configurations have the maximum length $2n+1$, but the reverse is not true. And the implication is not true if the lengths of the configurations are o($n$).

To give a more precise meaning to these questions, we now address the computation model in which we'll try to state them.

ECA can be simulated through 1-tape 2-symbols Turing machines if we adopt the following convention: a Turing machine T will be said to simulate an automaton if, starting with the number $n$ and the initial configuration (numbered 0) on its otherwise blank tape, it stops with the $n$ first configurations of the automaton written on its tape. How are we to recognize the different states? A simple convention makes that easy: let's say that the number representing a configuration will be written with all the symbols doubled. For instance, starting from the left to the right, a line containing two black cells followed by a blank one and then a black one (1101) will be represented on the tape by "11110011". This insures that the string "01" with 0 at an odd position will never appear and can be reserved for separating the different lines. Thus, at the end of the computation, the tape will start by "01" followed by the value of the first line (with figures doubled) then "01" separating the first line from the second one and so on…The tape will be by ended by "01". For example:
011101110011011111100110001
describes 3 lines. The first line with a black cell, the second line with a black cell, a blank one and then a black one, and the third line with cells black, black, white, black, white. This convention allows as well to specify the initial state of the ECA we want to simulate. At the beginning, the tape will contain

"01" then the number *n* of configurations to be computed (written with figures doubled) then "01" then the initial configuration (written with the same convention).

It's easy to see that this way of encoding ECA's through 1-tape 2-symbols Turing machines is not very efficient. The head will have to go back and forth permanently: back to read the three cells of the configuration *i*-1 that correspond to the current computed cell of the configuration *i*, forth to write the current cell, back again to read the next three cells of the configuration *i*-1, forth to write the next cell and so on. Since the configuration *i* has (2*i*+1) cells, the head will have to go back and forth (2*i*+1) times. During each trip, it will go through 2(2*i*-1) cells. With our coding that implies 4(2*i*-1)+4 moves. Thus it will need (2*i*+1)[4(2*i*-1)+4] steps for computing the configuration *i* from configuration *i*-1 (apart some details that doesn't change the argument). The total number of steps for computing the $n^{th}$ configuration is then $\Sigma(2i+1)[4(2i-1)+4]$ for *i*=1 to *n*. Since $\Sigma (2i+1)[4(2i-1)+4] = O(\Sigma i^2) = O(n^3)$, we compute ECA in $O(n^3)$. We could save space in allowing 3 symbols. That would avoid doubling each bit. But the gain would only be linear.

Is it possible to be more efficient? Yes, is suffices to use 2-tape 2-symbols Turing machines. When the computation starts, one tape contains the number of configurations that is to compute and the initial configuration, the second tape is blank. Now, the head of the first tape reads the initial configuration and the head of the second tape write the next configuration according to the update rules. It's easy to see that reading configuration *i*-1 and writing configuration *i* can be done in (2*i*+1) steps (here, no need to go back and forth). When configuration *i* is written the role of the heads is reversed. If the initial configuration has length 1, apart from a small subtlety to deal with the counter *n* allowing to stop after having written *n* configurations (which just adds a linear number of steps), the total number of steps when the machine halts is $\Sigma (2i+1)$ for *i*=1 to *n*: i.e. $O(n^2)$. A length of the initial configuration strictly greater than 1 will only have a constant multiplicative impact.

Can we save more time? Here the answer is no for unless an intermediate computation collapses, we have seen that the number of updating rules to perform is $n^2$-1. So we can say that in general the most efficient Turing machines for simulating ECA's can be chosen among 2-tape 2-symbols Turing machines computing in $O(n^2)$.

**Definition** (ECA *Turing machine*) *Let's denote by* $E_n$ *the* $n^{th}$ *state of the* ECA A. *A Turing machine* $T_A$ *will be called a* ECA *Turing machine representing* A *if:*
  - *For all n,* $T_A$ *computes* $E_n$ *on input n.* (*It's important to notice that this is the same Turing machine which on input n computes* $E_n$: $E_n$ *is uniformly computed by* $T_A$).
  - *during the computation, the* $T_A$ *tapes contain successively in an increasing order from i=1 to* n-1, *the configurations* $E_i$

In the following, we will abbreviate "for all *n*, T computes $E_n$ on input *n*" by "T computes every $E_n$".

A ECA Turing machine representing an automaton A is exactly a program simulating the behavior of the automaton.

We can now translate the question to decide if an automaton A is CIR or not in a question expressed in terms of ECA Turing machines representing A: Let A be a ECA and let $T_A$ be a ECA Turing machine representing A and running in time $O(n^2)$:

1) is it possible to find a Turing machine which on input *n* computes $E_n$ faster than $T_A$?

2) is it possible to find a Turing machine which on input *n* computes $E_n$ without computing the (*n*-1) previous $E_i$ (i.e. which is not a ECA Turing machine representing A)?

## 6. Tentative definitions

We'll consider first definitions in the framework of ECA Turing machines but, remembering that we seek a general definition, we shall come back to the case of functions *f(n)* from **N** to **N** to check is these definitions are robust enough.

Before giving our preferred definitions, it's worth presenting previous attempts linked to specific intuitions. These definitions, while intuitively appealing, don't work for reasons that preclude to use them as capturing correctly the concept of CIR.

### 6.1. Definition linked to the algorithmic complexity

**Tentative definition 1 (CIR)** *A* ECA *will be said* CIR *if and only if* $\exists c > 0, \forall n, K(E_n) > K(n) + c$

The intuition behind this definition is that the length of the states grows as *n* and thus that the ECA is not evolving through fixed states or states that oscillate or vanish. In a certain sense, the automaton keeps the memory of the number of iterations for reaching a given configuration. But this definition is too weak since it is respected by many automata that are obviously not CIR: for example, the automaton which, starting from a unique black cell, adds a black cell on each side of the previous configuration. The $n^{th}$ configuration contains 2*n*-1 contiguous black cells so $K(E_n)$ is of the order of magnitude of K(*n*) but this automaton is not CIR. Besides that, this definition, while a priori appealing for ECA is not applicable to function *f(n)* like predicates for example, since the only values that a predicate can take are 0 or 1. So the complexity of *f(n)* can't grow.

### 6.2. Definition linked to the logical depth

**Tentative definition 2 (CIR)** *A* ECA *will be said* CIR *if and only if* $\forall n, D(E_{n+1}) > D(E_n)$ *where* D, *the Bennett's logical depth, is understood as a measure of the content of computation.*

As we saw previously, the more a string is profound the more it needs time to be computed by its minimal (or near minimal) program. The intuition here is that the successive configurations of a CIR automaton should result from more and more computation. For a ECA satisfying this definition, it would be necessary to compute longer to get the $(n+1)^{th}$ configuration than to get the $n^{th}$ one. Its configurations would be deeper and deeper. The problem with this definition is that one can imagine that the behavior of a CIR automaton is such that even if in the average the configurations become deeper and deeper, it can happen that suddenly there is a fall in the successive depths. So the following definition is preferable.

**Tentative definition 3 (CIR)** *A* ECA *will be said* CIR *if and only if* $\forall n, D(E_n) = \Omega(n^2)$

We saw that to compute the $n^{th}$ state the ECA needs $n^2$ steps when each $k^{th}$ configuration contains 2*k*+1 cells. The intuition is that nothing is lost in the computation and that in the long range, about $n^2$ steps are necessary to compute the *n*th configuration even if occasionally the depth of one configuration can drop down to a lower value.

What is wrong with these definitions? It seems that they capture correctly the fact that the *n*th state needs a lot of computation in average to be produced and that it is not possible to get it quickly. Isn't it the very meaning of CIR? Actually not. Once again, the consideration of predicates is a good way to see the problem. The *n*th state of a predicate function will be either 0 or 1 and can neither be complex nor deep.

Considering the case of possibly CIR predicates shows that the real meaning of CIR is located inside the very succession of states not inside any single state. CIR is meaningful only regarding the way the different states are related each others. Definitions 2 and 3 mean that for each state, the time required to compute it from its minimal program must be long but this minimal program can be different for each state. On the contrary, what CIR means implies that there exists no general program which can compute for every *n* the *n*th state from n faster than the ECA. So that's a different meaning we need to capture. Nevertheless, this definition is interesting by itself and worthwhile to be explored. This will be done in another paper.

## 7. Preferred definitions

### 7.1. A first attempt

Let A be a ECA and let $T_A$ be a ECA Turing machine representing A and running in time $O(n^2)$, we saw that the question of deciding if a ECA A is CIR or not can be split into:
1) is it possible to find a Turing machine which computes every $E_n$ faster than $T_A$?
2) is it possible to find a Turing machine which on input *n* computes $E_n$ without computing the (*n*-1) previous $E_i$ (i.e. which is not a ECA Turing machine representing A)?

We now turn to the problem of giving a precise formulation of these questions. The first question will lead to the tentative definition 4 and the second question to the definition 5.

**Definition (*efficient* ECA *Turing machine*)** *We will say that a* ECA *Turing machine* $M_A^{eff}$ *representing* A *is an efficient* ECA *Turing machine representing* A *if no other* ECA *Turing machine representing* A *can compute the configuration* $E_n$ *faster than* $M_A^{eff}$. *Let* $T(M_A^{eff}(n))$ *the time for* $M_A$ *to compute* $E_n$. *More precisely, we will say that a* ECA *Turing machine* $M_A^{eff}$ *representing* A *is an efficient* ECA *Turing machine representing* A *if for any other* ECA *Turing machine* $M_A$ *representing* A: $T(M_A^{eff}(n)) = O(T(M_A(n)))$ *i.e. there are constants* $c > 0$, $n_0 > 0$ *such that* $\forall n > n_0$, $T(M_A^{eff}(n)) \leq cT(M_A(n))$ *where* $T(M_A(n))$ *is the time for* $M_A$ *to compute* $E_n$.

Of course when $T(M_A(n))=O(n^2)$, it is always possible to design a ECA Turing machine representing A and computing in time greater than $O(n^2)$, if for example, the related program is not efficient. But, as we saw previously, it is also possible to have $T(M_A(n)) = O(n)$ for simpler automata such that rules 4, 12, 36 and many similar others[32]. For these automata, the number of cells that change between two successive configurations is bounded by a constant and so is the time to compute the $(n+1)^{th}$ configuration from the $n^{th}$ configuration. As we said previously, we don't want to call CIR these automata. So we will demand that $T(M_A(n)) = O(n^2)$ (i.e. automata of classes 3 and 4) and of course, we will exclude class 3. This leads to the following definition.

---

[32] $T(M_A(n)) = O(n)$ or $T(M_A(n)) = O(n^2)$. One could think of an intermediate situation with an automaton for which $O(n^2) > T(M_A(n)) > O(n)$. For example $T(M_A(n)) = O(n \log n)$. Interestingly enough, none of the 256 ECA has this property.

**Tentative Definition 4 (CIR)** *A ECA A will be said CIR if and only if no Turing machine computing every $E_n$ computes the configuration $E_n$ in a number of steps less than $O(n^2)$.*

The intuition behind this definition is that it is impossible to compute the state $E_n$ faster that the automaton itself. There is no way to predict the result of the computation done by the automaton because in order to know what is the state $E_n$, whatever general program is used, it will need as much time as running the automaton itself. What's wrong with this definition? We have just defined CIR as the fact that the most efficient program to compute $E_n$ can't run faster than the simulation. That is of course a first step (linked to the first part of the question above) but that is not enough to capture the whole intuition about CIR. The intuition that we want to preserve is that in order to know the state of the system after *n* steps it is necessary to follow approximately the same path if not exactly the same. The main difficulty here is to define exactly what is meant by "approximately the same path".

## 7.2. The final definition

**Definition (*approximation of a ECA Turing machine*)** *Let $M_A^{eff}$ be an efficient ECA Turing machine representing A and computing in $T(M_A^{eff}(n))$. A Turing Machine T will be said to be an approximation of a ECA Turing machine representing A if and only if there is a function F such that $F(n)=o(T(M_A^{eff}(n))/n)$ and a Turing machine P[33] such that:*

*(i) on input n, T computes a result $r_n$ and halts.*
*(ii) during the computation, the T tape contains as a substring successively in an increasing order from i=1 to n-1, data $r_i$ from which the Turing machine P computes $E_i$ in a number of steps F(i).*

Intuitively, an approximation of a ECA Turing machine is a Turing machine doing a computation that is near the computation made by an ECA Turing machine. We are going to show below how it is possible to build a ECA Turing machine from any approximation of a ECA Turing machine. The factor $1/n$ in $o(T(M_A(n))/n)$ takes into account the fact that there are *n* necessary steps to compute $E_n$ with a ECA Turing machine and that we want the computation time of P between $r_i$ and $E_i$ to be much shorter than the average of the computation time between $E_{i-1}$ and $E_i$.

**Definition 5 (CIR ECA)** *A ECA A will be said CIR if and only if any Turing machine computing every $E_n$ is an approximation of a ECA Turing machine representing A.*

**Theorem 7.2.** *If a ECA A is CIR then no Turing machine computing every $E_n$ can compute faster than an efficient ECA-Turing machine representing A. More precisely, if M is a Turing machine computing every $E_n$ then $T(M_A^{eff}(n)) = O(T(M(n)))$.*

**Proof:** We shall start by proving the following result.

---

[33] Here again, it's important to notice that this is the same Turing machine P which on input $r_i$ computes *f(i)*. That's a uniform computation.

**Lemma** *Given any* M *approximation of a* ECA *Turing machine representing A, there exists a* ECA *Turing machine* M' *representing* A (*we'll call it the daughter of* M) *computing in a time* $T(M'(n)) \backsim T(M(n))$.

**Proof**: Since M is an approximation of a ECA Turing machine representing A, there are a Turing machine P and a function F associated as mentioned in the definition. Let's consider the Turing machine M' which does exactly the same computation than M but for each $i$, when $r_i$ is written on its tape, which computes $E_i$ through P from $r_i$ in a time $F(i)$, writes $E_i$ on its tape and resumes the computation at the exact point where it left M computation. It's clear that M' is a ECA Turing machine representing A. From the fact that $F(n)=o(T(M_A(n))/n)$ the additional time will be at most $o(T(M_A(n)))$. Hence, M' will compute in $T(M'(n))= T(M(n)) + o(T(M_A(n)))$. Since M' is a ECA Turing machine representing A, $T(M'(n)) \geq TM_A(n)$. Hence $T(M(n)) + o(T(M_A(n))) \geq T(M_A(n))$. From that, it follows that:

$$\lim_{n \to \infty} \frac{T(M'(n))}{T(M(n))} = \lim_{n \to \infty} \left(1 + \frac{o(T(M_A(n)))}{T(M(n))}\right) = 1$$

Therefore, $T(M'(n)) \backsim T(M(n))$. M and its daughter M' compute in the same time.

We can now prove the theorem.
If A is CIR, a Turing machine M computing every $E_n$ is an approximation of a ECA Turing machine representing A. From the lemma, M', the daughter of M (which computes in the same time than M) is a ECA Turing machine representing A. So $T(M_A^{eff}(n)) = O(T(M'(n)))$ and since $T(M'(n)) \backsim T(M(n))$ we get $T(M_A^{eff}(n)) = O(T(M(n)))$.

Through theorem 7.2, it is clear that if a ECA A is CIR, it will be impossible to compute its $n^{th}$ configuration faster than A itself (i.e. an efficient ECA Turing machine representing A). Hence A will satisfy automatically the tentative definition 4. It will also be necessary to follow a path similar to the path followed by A. That's exactly the meaning (now fully formalized) of Wolfram's claim.

*Remark: Previously, we explicitly discarded* ECA *of class 2 (with simulation time* $O(n)$*) because we didn't want to accept that any of them be CIR. In definition 5, we don't suppose that the efficient* ECA *Turing machine representing a CIR* ECA *must computes in* $O(n^2)$*. So, it seems that we open the door to find a CIR* ECA A *such that an efficient* ECA *Turing machine representing* A *computes in* $O(n)$*, provided that any Turing machine computing every* $E_n$ *is an approximation of a* ECA *Turing machine representing* A*. Actually, it's easy to see that none of the class 2* ECA *has this property since for all of them it is possible to compute directly the* $n^{th}$ *configuration without having to compute all the previous ones.*

## 8. Generalization

We now leave the ECA framework and switch to the generalization of the definition to any function from **N** to **N**. We mimic the same process of definitions for functions than for ECA. In the following *f* will be a function from **N** to **N**.

**Definition (E-*Turing machine*)** *A Turing machine* $T_f$ *will be called a* E-*Turing machine representing f if:*
- $T_f$ *computes every f(n).*
- *during the computation, the* $T_f$ *tape contains as a substring successively in an increasing order from i=1 to n-1, the values f(i).*

A E-Turing machine representing a function *f* is a program enumerating the function *f* through the computation of the successive values *f(i)*. It is the equivalent for a given function of what a ECA Turing machine simulating the ECA through the enumeration of all its successive states is for the ECA.

**Definition (*Efficient* E-*Turing machine*)** *We will say that a* E-*Turing machine* $M_f^{eff}$ *representing f is an efficient* E-*Turing machine representing f if no other* E-*Turing machine representing f can compute f(n) faster than* $M_f$. *Let* $T(M_f^{eff}(n))$ *the time for* $M_f$ *to compute f(n). More precisely, we will say that a* E-*Turing machine* $M_f^{eff}$ *representing f is an efficient* E-*Turing machine representing f if for any other* E-*Turing machine* $M_f$ *representing f:* $T(M_f^{eff}(n)) = O(T(M_f(n)))$ *i.e. there are constants* $c > 0, n_0 > 0$ *such that* $\forall n > n_0, T(M_f^{eff}(n)) \leq cT(M_f(n))$

**Definition (*approximation of a* E-*Turing machine*)** *Let* $M_f^{eff}$ *be an efficient* E-*Turing machine representing a function f. For every input n,* $M_f^{eff}$ *computes f(n) and halts in a time* $T(M_f^{eff}(n))$. *A Turing Machine* T *will be said to be an approximation of a* E-*Turing machine representing f if and only if there is a function* F *such that* $F(n)=o(T(M_f^{eff}(n))/n)$ *and a Turing machine* P *such that:*
*(i) on input n,* T *computes a result* $r_n$ *and halts.*
*(ii) during the computation, the* T *tape contains as a substring successively in an increasing order from i=1 to n-1, data* $r_i$ *from which* P *computes f(i) in a number of steps* F(i).

**Definition 5 (*CIR function*)** *A function f(n) from* **N** *to* **N** *will be said* CIR *if and only if any Turing machine computing every f(n) is an approximation of a* E-*Turing machine representing f.*

**Theorem 8.1.** *If a function f is* CIR *then no Turing machine computing every f(n) can compute faster than an efficient* E-*Turing machine representing f. More precisely, if* M *is a Turing machine computing every f(n) then* $T(M_f^{eff}(n)) = O(T(M(n)))$.

**Proof:** The proof is identical to the proof of theorem 7.2.

## 9. Conclusion

We derived a robust definition for the computational irreducibility and we proved that, through a robust definition of what means "to be unable to compute the state *n* without having to follow the same path than the computation simulating the automaton or the function", this implies genuinely, as intuitively expected, that if the behavior of an object is computationally irreducible, no computation of its $n^{th}$ state can be faster than the simulation itself.
For CIR automata, functions or processes, there is no short-cut allowing to predict the $n^{th}$ state faster than waiting till all the steps are done. In this sense, these objects are unpredictable.

An open problem is now to prove that explicit objects are really CIR. Possible candidates are:

- $F(n) = 2^n$ (in base 10)
- $F(n) = n!$
- F(n) = the $n^{th}$ prime number
- $x_n = 4x_{n-1}(1-x_{n-1})$ (logistic map for $x_0 \in [0,1]$ and having a limited number of digits)
- Rule 30
- Rule 110
- ………

This opens the way for philosophical discussions and applications. Assume for example that the process leading from inanimate matter to life, or from neurons to consciousness be CIR. In this case, there could be no way to "understand" what is life or consciousness since understanding a phenomenon is being able to predict the final state from the initial conditions, or at least to anticipate intuitively what is going to happen. For a CIR process this is impossible[34]. CIR could also be a key for explaining emergent phenomena. These points will be studied elsewhere.

The concept of approximation can be extended so as to give a classification of computable functions with similar properties in terms of irreducibility, algorithmic complexity and logical depth. This will be presented in the part II of this paper (forthcoming).

---

[34] See a more extensive discussion of this point in [Zwirn, 2006].